\documentclass[a4paper, 11pt]{article}
\usepackage{amsmath}
\usepackage{graphicx}
\usepackage{latexsym}

\def\b{\begin{eqnarray}}
\def\e{\end{eqnarray}}
\def\n{\noindent}

\begin{document}

\begin{center}

{\LARGE\textbf{Hamiltonian formulation and integrability of a
complex symmetric nonlinear system
\\}} \vspace {10mm} \vspace{1mm} \noindent

{\large \bf Rossen Ivanov$^\ast$}\footnote{On leave from the
Institute for Nuclear Research and Nuclear Energy, Bulgarian Academy
of Sciences, Sofia, Bulgaria.} \vskip1cm \n \hskip-.3cm
\begin{tabular}{c}
\hskip-1cm $\phantom{R^R}${\it School of Mathematics, Trinity
College,}
\\ {\it Dublin 2, Ireland} \\ {\it Tel:  + 353 - 1 - 608 2898 }\\{\it  Fax:  + 353 - 1- 608 2282} \\
\\ {\it $^\ast$e-mail: ivanovr@tcd.ie} \\
\\
\hskip-.8cm
\end{tabular}
\vskip1cm
\end{center}

\vskip1cm

\begin{abstract}
The integrability of a complex generalisation of the 'elegant'
system, proposed by D. Fairlie \cite{DF} and its relation to the
Nahm equation and the Manakov top is discussed.



\end{abstract}

\newpage

\section{A complex nonlinear system with symmetry}

The Euler top is probably the best known example of a nonlinear
integrable system. Generalizations of this system are under intense
investigation for many years. In general, the (generalized) top
equations contain quadratic nonlinearity. A problem of the form
$dY/dt=Y^2$, where $Y$ is an $n\times n$ symmetric matrix with
vanishing diagonal elements is analyzed in the work of Fairlie
\cite{DF} and leads to the following $n$-dimensional nonlinear
system:

\begin{equation}
\frac{du_{j}}{dt}=\prod _{k\neq j}^n u_{k}, \qquad j=1,\ldots,n.
\label{eq:01}
\end{equation}
In \cite{DF} this system is explicitly integrated. The system
(\ref{eq:01}) for $n=4$ appears also in the classification of a
class of double-elliptic systems \cite{BGOR}. Other higher
dimensional integrable tops have been recently found in
\cite{FU1,FU2,FU3}.

The problem of generalizations of the standard Hamiltonian dynamics
to complex dynamical variables often contains many interesting
aspects, e.g. \cite{7,7a}. In particular, an integrable
complexification of (\ref{eq:01}) for $n=3$ and the related Halphen
system are considered in \cite{2}. A natural generalization of these
systems is the following system in $n$ complex dimensions:

\begin{equation}
\frac{d\bar{u}_{j}}{dt}=\prod _{k\neq j}^n u_{k}, \qquad
j=1,\ldots,n, \label{eq:02}
\end{equation}
where the bar stands for complex conjugation. The system
(\ref{eq:02}) is also integrable. Indeed, it possesses the following
independent algebraic constants of the motion:
\begin{eqnarray}
H &\equiv &iI_{1}=\prod _{k=1}^n \bar{u}_{k}-
\prod _{k=1}^n u_{k},  \label{eq:03} \\
I_{k} &=&|u_{1}|^{2}-|u_{k}|^{2},\;k=2,...n.  \label{eq:04}
\end{eqnarray}
There are $n$ real integrals, $I_{1}$, ..., $I_{n}$, but they are
sufficient for the integration of the system.

The system (\ref{eq:02}) can be realised as a reduction of the
following Hamiltonian system with $\mathbf {Z_2}$ symmetry:
\begin{eqnarray}
H (p,q)&\equiv &\prod _{k=1}^n p_{k}- \prod _{k=1}^n q_{k},
\nonumber\\ {\frac{d p_j}{dt}}&=& -{\frac{\partial H}{\partial
q_{j}}}= \prod _{k\neq j}^n q_{k}  \label{eq:p} \\
{\frac{d q_j}{dt}}&=&{\frac{\partial H}{\partial p_{j}}}=\prod
_{k\neq j}^n p_{k}\label{eq:q}
\end{eqnarray}

\n Considering $q_k\equiv u_{k}$ as coordinate variables and $p_k
\equiv \bar{u}_{k}$ as momentum variables, the system (\ref{eq:02})
admits a Hamiltonian formulation with a Hamiltonian $H$ given by
(\ref{eq:03}). The equations of motion (\ref{eq:p}) and (\ref{eq:q})
then coincide with (\ref{eq:02}). Note that in the real case
(\ref{eq:01}) $H\equiv 0$.

Now we are able to integrate explicitly the system (\ref{eq:02}).
Let $u_{k}=e^{i\phi _{k}}r_{k}$ where $r_{k}$, $\phi _{k}$ are real
functions and $\phi = \phi _{1}+\phi _{2}+...+\phi _{n}$. Then
\[
u_{k}d\overline{u}_{k}+\overline{u}_{k}du_{k}=rdt
\]
where $r=\prod _{k=1}^n u_{k}+\prod _{k=1}^n \overline{u}_{k}$ is a
real function. Let
\begin{equation}
rdt=d\tau.   \label{eq:2}
\end{equation}
\n Then
\begin{equation}\label{eq:r}
dr_{k}^{2}=rdt
\end{equation}
\n and
\begin{equation}
r_{k}^{2}=\tau +c_{k}  \label{eq:3}
\end{equation}
for some real constants $c_{k}$. It is obvious that
\begin{eqnarray}
I_{1} &=&-2\left( \prod _{k=1}^n r_{k}\right) \sin \phi
   \label{eq:i33} \\
r &=&2\left( \prod _{k=1}^n r_{k}\right) \cos \phi \label{eq:i34}
\end{eqnarray}
and therefore from (\ref{eq:2}), (\ref{eq:3}), (\ref{eq:i33}) and
(\ref{eq:i34}) we obtain
\begin{equation}
\left( {\frac{d\tau }{dt}}\right) ^{2}=4\left( \prod _{k=1}^n
r_{k}\right) ^{2}-I_{1}^{2}= 4\left( \prod _{k=1}^n (\tau
+c_{k})\right) -I_{1}^{2}.  \label{eq:4}
\end{equation}
i.e. formally
\begin{equation}
t=\int {\frac{d\tau }{\sqrt{4\left( \prod _{k=1}^n (\tau
+c_{k})\right) -I_{1}^{2}}}}  \label{eq:5}
\end{equation}
For a specific choice of the constants $c_{k}$ and $I_{1}$, $\tau $
can be expressed via $t$ through a hyperellyptic function.
Furthermore
\[
{\frac{d}{dt}}\left( {\frac{u_{k}}{\bar{u}_{k}}}\right)
=i{\frac{I_{1}}{\bar{u}_{k}^{2}}},
\]
i.e.
\begin{equation}\label{eq:phik}
{\frac{d\phi _{k}}{dt}}={\frac{I_{1}}{2r_{k}^{2}}},
\end{equation}
and from (\ref{eq:3}) and (\ref{eq:4}) we have formally
\begin{equation}
\phi _{k}(\tau )={\frac{I_{1}}{2}}\int {\frac{d\tau }{(\tau +c_{k})\sqrt{%
4\left( \prod _{k=1}^n (\tau +c_{k})\right) -I_{1}^{2}%
}}}.  \label{eq:6}
\end{equation}
The integrals (\ref{eq:5}) and (\ref{eq:6}) can be expressed in
terms of hyperellyptic functions.

In order to better understand the underlying symmetry of the system
let us rewrite it in terms of real variables as follows. The
summation over $k$ of (\ref{eq:phik}) gives

\begin{equation}\label{eq:phi}
{\frac{d\phi}{dt}}={\frac{I_{1}}{2}\sum _{k=1}^n
\frac{1}{r_{k}^{2}}},
\end{equation}
\n which, due to (\ref{eq:i33}) is equivalent to

\begin{equation}\label{eq:phi1}
{\frac{d\phi}{dt}}=-\sin \phi \left( \prod _{k=1}^n r_{k}\right)
\sum _{k=1}^n \frac{1}{r_{k}^{2}}.
\end{equation}

\n From (\ref{eq:r}) and (\ref{eq:i34}) we obtain the following set
of equations:

\begin{equation}\label{eq:rkf}
{\frac{d r_k}{dt}}=\cos \phi  \prod _{i\neq k}^n r_{i}.
\end{equation}

\n Now it is clear that (\ref{eq:02}) is equivalent to the $n+1$
dimensional real system (\ref{eq:phi1}), (\ref{eq:rkf}). The
reduction to the 'real' version (\ref{eq:01}) can be achieved as
follows. Let us choose for $\phi$ the initial value $\phi(0)=2K\pi$
for some arbitrary integer $K$. Then the integral $I_1=0$, according
to (\ref{eq:i33}), and consequently ${\frac{d\phi}{dt}}=0$,
according to (\ref{eq:phi}). Thus the solution in this case is
$\phi(t)=2K\pi$ and (\ref{eq:rkf}) simply coincides with
(\ref{eq:01}).

\section{Example: $su(3)$ Nahm top ($n=3$)}

It is well known that the one-dimensional reductions of the
Self-dual Yang Mills (SDYM) equations in four dimensions yield
various generalizations of some classical systems as the
Euler-Arnold-Manakov and Nahm equations \cite{1,Nahm}. In \cite{2}
the SDYM equations in seven dimensions have been investigated, in
the case when their invariance group is the $\textbf{G}_{2}$
subgroup of the rotation group $SO(7)$. One of the six-dimensional
reductions of these equations leads to (\ref{eq:02}) with $n=3$:
\begin{equation}
{\frac{d\bar{u}_{1}}{dt}}=u_{2}u_{3}\quad \text{and cyclic
permutations}, \label{top1}
\end{equation}
\n In \cite{2} the question about the existence of a Lax pair for
the system (\ref{top1}) is posed, but the Lax pair is not found.
Although this system is in principle known for some time, since it
is an interesting one here we pay some more attention to it.

In physical terms, the system (\ref{top1}) describes membrane
instantons in three complex dimensions \cite{FL,CFZ}. It can also be
seen as a ''stationary point'' for the $3$--wave equations with a
''blow-up'' instability (i.e. ${\frac{\partial u_{i}}{\partial
x}}=0$):
\[
{\frac{\partial u_{1}}{\partial t}}+v_{1}{\frac{\partial
u_{1}}{\partial x}} =i\epsilon \overline{u}_{2}\overline{u}_{3},
\]
where $v_{k}$ and $\epsilon $ are real constants \cite{4,4a,8}. The
equations (\ref{top1}) possess a complete symmetry and in this
respect are similar to the Kaup equations, where the derivatives are
over the three space variables \cite{K}.

Firstly, one can show that (\ref{top1}) is related to the well known
Nahm equation. Using the matrices

\begin{eqnarray}
T_1&=&\left(
\begin{array}{ccc}
0 & 0 & 0 \\
0 & 0 & -u_{1} \\
0 & \bar{u}_{1} & 0
\end{array}
\right), \quad  T_2=\left(
\begin{array}{ccc}
0 & 0 & \bar{u}_{2} \\
0 & 0 & 0 \\
-u_{2} & 0 & 0
\end{array}
\right) , \nonumber \\  T_3&=&\left(
\begin{array}{ccc}
0 & -u_{3} & 0 \\
\bar{u}_{3} & 0 & 0 \\
0 & 0 & 0
\end{array}
\right) \label{matrices}
\end{eqnarray}

\n the system (\ref{top1}) can be written in the usual form for the
Nahm equation

\begin{equation}
{\frac{dT_i}{dt}}=\frac{1}{2}\varepsilon_{ijk}[T_j,T_k].
\label{top2}
\end{equation}

\n Equation (\ref{top2}) was introduced initially in relation to
$SU(2)$ monopoles on $\bf{R^3}$ \cite{Nahm}. However, the Nahm
matrices, $T_i$ in (\ref{top2}) can take values in various simple
Lie algebras (not necessarily $su(2)$). This is explained in details
in \cite{3}. Since the matrices (\ref{matrices}) are anti-Hermitian,
i.e. take values in $su(3)$, we will call (\ref{top1}) an $su(3)$
Nahm top. Now it is straightforward to represent (\ref{top2}) in the
Lax form $\frac{dL}{dt}=[L,M]$ (e.g. see \cite{NH,NMM}) with \b
\label{LNahm} L(\zeta)&=&T_1+iT_2-2\zeta i T_3+\zeta^2(T_1-iT_2) \\
\label{MNahm} M(\zeta)&=&-iT_3+\zeta(T_1-iT_2).\e

\n The computation of $\det(L(\zeta)-\lambda \mathbf{1})$
immediately gives the integrals of motion (\ref{eq:03}),
(\ref{eq:04}). For another form of the Lax pair for the Nahm
equation see also \cite{CFZ}.

The system (\ref{top1}) can also be seen as a Manakov top \cite{5}
with a complex inertia tensor. Indeed, consider the following Lax
pair:

\b \label{LMan} L(\zeta)&=&[J^2,Q]-\zeta J^2
\\ \label{MMan} M(\zeta)&=&[J,Q]-\zeta J,\e
where the cubic root of unity, $\omega =e^{2\pi i/3}$ is used to
define the inertia tensor
\begin{equation}
J=\frac{1}{\omega(\omega-1)}J_0, \qquad J_0 =\text{diag}\,(\omega
,\omega ^{2},1); \label{top4}
\end{equation}
the matrix $Q$ is defined as $[J_0,Q]=-(T_1+T_2+T_3)$ where where
$T_k$ are the matrices (\ref{matrices}), i.e.
$Q=-\mathrm{ad}_{J_{0}} ^{-1} (T_1+T_2+T_3)$.
Explicitly, the operators $L$ (\ref{LMan}), and $M$ (\ref{MMan}) are
(the irrelevant overall factor $[\omega(\omega-1)]^{-2}$ in $L$ is
omitted)
\b \label{eL} L(\zeta)&=&\sum_{k=1}^{3}\omega^kT_k-\zeta J_0 ^2,\\
\label{eM} M(\zeta)&=&-\frac{1}{\omega (\omega
-1)}\Big(\sum_{k=1}^{3}T_k+\zeta J_0\Big). \e

\n  The characteristic polynomial of $L(\zeta)$, (\ref{eL}) is

\begin{equation}
\det L(\zeta)=-\zeta^3-\zeta \Big(\omega |u_1|^2+\omega ^2 |u_2|^2+
|u_3|^2 \Big)+(\bar{u_1}\bar{u_2}\bar{u_3}-u_1u_2u_3)
\end{equation}
Its coefficients can be split into real and imaginary parts to give
the integrals of motion (\ref{eq:03}), (\ref{eq:04}).

The representation of the system (\ref{top1}) as a Manakov top
allows immediately the classical $r$-matrix to be written
\cite{FT87,KS}. Defining a Poisson bracket
\begin{equation}
\{A \stackrel{\bigotimes} {,}  B \}\equiv
\sum_{k=1}^{3}\Big(\frac{\partial A}{\partial u_k}\otimes
\frac{\partial B}{\partial \bar{u}_k}-\frac{\partial A}{\partial
\bar{u}_k}\otimes\frac{\partial B}{\partial u_k}\Big),
\end{equation}

\n it is straightforward to check, that

\begin{equation}
\{L(\zeta) \stackrel{\bigotimes} {,}
L(\mu)\}=[r(\zeta-\mu),L(\zeta)\otimes \mathbf{1}+\mathbf{1}\otimes
L(\mu)],
\end{equation}

\n where $L$ is given by (\ref{eL}), and the classical $r$ matrix is

\begin{equation}
r(\zeta)=-\frac{P}{\omega (\omega -1)\zeta}.
\end{equation}

\n Here $P$ is the permutation matrix in $\mathrm{C^3}\otimes
\mathrm{C^3}$, i.e.

\begin{equation}
P(x\otimes y)=y\otimes x, \qquad x,y\in \mathrm{C^3}.
\end{equation}






\section{Example: $so(4)$ Nahm top ($n=4$)}

As a second example we consider the system (\ref{eq:01}) with $n=4$.
It is known that with a change of variables \cite{BGOR} it can be
represented as two $so(3)$ Nahm tops. Since $so(4)=so(3)\oplus
so(3)$ it is natural to expect that (\ref{eq:01}) with $n=4$ is an
$so(4)$ Nahm top. Indeed, using the $so(4)$ parametrization given in
\cite{6} we can define the following $so(4)$ Nahm matrices:

\begin{eqnarray}
T_1&=&\left(
\begin{array}{cccc}
0 & 0 & 0 &u_1 u_4\\
0 & 0 & -u_{2}u_3&0 \\
0 & u_2 u_3 & 0 & 0 \\
-u_1 u_4 & 0 & 0& 0
\end{array}
\right), \nonumber\\ T_2&=&\left(
\begin{array}{cccc}
0 & 0 & u_1 u_3&0\\
0 & 0 &0 & u_{2}u_4 \\
-u_1 u_3 & 0 & 0 & 0 \\
0 & -u_2 u_4 & 0& 0
\end{array}
\right), \nonumber \\  T_3&=&\left(
\begin{array}{cccc}
0 & -u_3 u_4 & 0 &0\\
u_3 u_4 & 0 &0 & 0 \\
0 & 0 & 0 & u_1 u_2 \\
0 & 0 & -u_1 u_2 & 0
\end{array}
\right).  \label{so4matrices}
\end{eqnarray}

\n With the help of (\ref{so4matrices}), one can write (\ref{eq:01})
explicitly in the canonical form (\ref{top2}). Now the Lax pair
(\ref{LNahm}), (\ref{MNahm}) follows automatically. We also note
that the construction

\b \label{eL1} L&=&\sum_{k=1}^{3}\omega^k T_k,\\
\label{eM1} M&=&-\frac{1}{\omega (\omega -1)}\sum_{k=1}^{3}T_k \e

\n also provides a Lax pair for this system, indeed, without a
spectral parameter.

The Lie group interpretation, the Lax pairs etc. for the system
(\ref{eq:02}) with $n\geq 4$ is a very interesting problem, which
deserves a further investigation.

\section*{Acknowledgements}
The author is grateful to V. Gerdjikov, G. Grahovski and N. Kostov
for many valuable discussions. Funding from Science Foundation
Ireland, Grant 04/BR6/M0042 is gratefully acknowledged.


\begin{thebibliography}{99}
\bibitem{DF}  D. Fairlie, Phys. Lett. A 119 (1987) 438.

\bibitem{BGOR}  H. Braden, A. Gorsky, A. Odesskii, V. Rubtsov, Nucl. Phys. B 633 (2002)
414; hep-th/0111066.

\bibitem{FU1}  D. Fairlie, T. Ueno,  Phys. Lett. A 248 (1998) 132.

\bibitem{FU2}  T. Ueno,  Phys. Lett. A 245 (1998) 373.

\bibitem{FU3}  D. Fairlie, T. Ueno,  J. Phys. A 31 (1998) 7785.

\bibitem{7}  V.S. Gerdjikov, A. Kyuldjiev, G. Marmo, G. Vilasi, Eur.
Phys. J. B 29 (2002) 177.

\bibitem{7a}  V.S. Gerdjikov, A. Kyuldjiev, G. Marmo, G. Vilasi, Eur. Phys. J. B
38 (2004) 635.

\bibitem{2}  K. Sfetsos, Nucl. Phys. B 629 (2002) 417; hep-th/0112117.

\bibitem{1}  S. Chakravarty, M.J. Ablowitz, P.A. Clarkson, Phys. Rev.
Lett. 65 (1990) 1085.

\bibitem{Nahm} W. Nahm,  The construction of all self-dual multimonopoles by the ADHM
method, in: Monopoles in Quantum Field Theory, N.S Craigie, P.
Goddard and W. Nahm (eds.), World Scientific, 1982, pp.87-95.

\bibitem{FL}  E.G. Floratos, G.K. Leontaris, Phys. Lett. B 545 (2002)
190.

\bibitem{CFZ} T. Curtright, D. Fairlie, C. Zachos, Phys. Lett. B 405 (1997)
37.

\bibitem{4}  V.E. Zakharov, S.V. Manakov, S.P. Novikov, L.P. Pitaevskii,
Theory of solitons: the inverse scattering method, Plenum, N.Y.,
1984.

\bibitem{4a}  V.E. Zakharov, S.V. Manakov, Exact theory of resonant interaction of wave
packets in nonlinear media, INF preprint 74-41, Novosibirsk (1975)
(In Russian).

\bibitem{8}  E.B. Gledzer, F.B. Drajanskii, A.M. Obukhov, Systems
of hydrodynamic type, Nauka, Moscow, 1981 (In Russian).

\bibitem{K}  D.J. Kaup, J. Math. Phys. 22 (1981) 1176.

\bibitem{3}  T. Brzezinski, H. Merabet, Czechoslovak J. Phys. 47 (1997)
1101; hep-th/9709070.

\bibitem{NH}  N.J. Hitchin, Commun. Math. Phys. 89 (1983) 145.

\bibitem{NMM}  N.J. Hitchin, N.S. Manton, M.K. Murray, Nonlinearity 8 (1995) 661.

\bibitem{5}  S.V. Manakov, Funct. Anal. Pril. 10 (1976)  93 (In
Russian).

\bibitem{FT87} L.D. Faddeev, L.A. Takhtajan, Hamiltonian methods in the theory of
solitons, Springer-Verlag, Berlin, 1987.

\bibitem{KS} P.P. Kulish and E.K. Sklyanin, Zap. Nauchn. Sem. LOMI 77 (1978) 134
(In Russian); J. Sov. Math. 22 (1983) 1627.

\bibitem{6}  L.A. Bordag, A.B. Yanovski, J. Phys. A: Math. Gen.
29 (1996) 5575.


\end{thebibliography}
\end{document}